\documentclass[aip,pop,amsmath,amssymb,preprint]{revtex4-1} 

\usepackage{graphicx}
\usepackage{dcolumn}
\usepackage{bm}
\usepackage[colorlinks=true,citecolor=blue,linkcolor=blue,urlcolor=blue]{hyperref}

\draft 

\begin{document}


\title{Observation of Toroidal Alfv\'{e}n Eigenmodes during Minor Disruptions in Ohmic Plasmas} 



\author{Yangqing Liu}
\affiliation{Department of Engineering Physics, Tsinghua University, Beijing 100084, China}
\author{Yi Tan}
\email[]{tanyi@sunist.org}
\affiliation{Department of Engineering Physics, Tsinghua University, Beijing 100084, China}
\author{Zhe Gao}
\affiliation{Department of Engineering Physics, Tsinghua University, Beijing 100084, China}
\author{Yuhong Xu}
\affiliation{Department of Engineering Physics, Tsinghua University, Beijing 100084, China}
\affiliation{Southwestern Institute of Physics, Chengdu 610041, China}
\author{Youjun Hu}
\affiliation{Institute of Plasma Physics, Chinese Academy of Sciences, Hefei, Anhui 230031, China}
\author{Song Chai}
\affiliation{Department of Engineering Physics, Tsinghua University, Beijing 100084, China}
\author{Yanzheng Jiang}
\affiliation{Department of Engineering Physics, Tsinghua University, Beijing 100084, China}
\author{Rui Ke}
\affiliation{Department of Engineering Physics, Tsinghua University, Beijing 100084, China}
\affiliation{Southwestern Institute of Physics, Chengdu 610041, China}
\author{Heng Zhong}
\affiliation{Department of Engineering Physics, Tsinghua University, Beijing 100084, China}
\author{Wenhao Wang}
\affiliation{Department of Engineering Physics, Tsinghua University, Beijing 100084, China}


\date{\today}

\begin{abstract}
  Toroidal Alfv\'{e}n eigenmodes (TAEs) excited in purely ohmically heated plasmas without any auxiliary heating have been identified for the first time in the SUNIST spherical tokamak. The TAE modes are observed during minor disruptions and have a frequency range of 150-500\,kHz. The mode structure analysis indicates the existence of both $m/n=-3/-1$ and $-4/-1$ harmonics, propagating in the electron diamagnetic direction in the laboratory frame of reference. These TAEs appear simultaneously with the generation of runaway electrons in the current quench phase, accompanying with the density sweeping during the minor disruption. Possible driving mechanisms and potential applications of these TAEs are discussed.
\end{abstract}

\pacs{52.35.Bj, 52.55.Fa, 52.55.Pi}

\maketitle 


Toroidal Alfv\'{e}n eigenmodes (TAEs) driven by fast particles are of particular importance for future burning plasma devices, where energetic particles will be abundantly produced by fusion reactions.\cite{Fasoli-2007-264-Chapter} As shown in experiments on the tokamaks TFTR\cite{Wong-1991-1874-Excitation} and D-IIID,\cite{Heidbrink-1991-1635-Investigation} TAEs are the most dangerous from the point of view of fast particle redistribution and losses. Furthermore, TAEs were first predicted theoretically to be excited by energetic particles,\cite{Cheng-1985-21-High} extensive efforts for exciting TAEs have been made in tokamak plasmas in order to understand the mechanisms. In present tokamak experiments, the destabilization of TAEs has been widely investigated in different devices using super-Alfv\'{e}nic ions generated by auxiliary heating systems such as neutral beam injection and ion cyclotron resonant heating\cite{Wong-1999-1-review,Aliarshad-1995-715-Observation} and active excitation by saddle coils.\cite{Fasoli-1995-645-Direct} As we know, the excitation of TAEs depends on the energy of the particles rather than the mass, so that energetic electrons can also drive these unstable TAEs. Fast electron driven TAEs were first seen on Compass-D with electron cyclotron heating only\cite{Valovic-2000-1569-Quasi} and on Alcator C-mod with lower hybrid current drive during the plasma current rising.\cite{Snipes-2008-72001-Fast} Runaway electrons (REs) are also a source of energetic particles and are naturally supposed to be able to excite Alfv\'{e}n eigenmodes such as TAEs and $\beta$-induced Alfv\'{e}n eigenmodes (BAEs). Recently, energetic electrons or REs are found to play a role in the activities of BAEs.\cite{Chen-2010-185004-Beta,Xu-2013-65002-Experimental,Liu-2015-65007-Observation} However, the TAE related to REs are rarely reported. Although Alfv\'{e}-type oscillations excited by REs have been observed in the ohmic regime of TUMAN-3M tokamak,\cite{Lebedev-2016-P5-Alfven} they are not TAEs. Magnetic fluctuations in the frequency range $f \simeq 60-260$\,kHz during disruptions without runaway plateaus have been observed in TEXTOR tokamak. \cite{Zeng-2013-235003-Experimental} Later theoretical studies suggested that such magnetic fluctuations may be TAE driven by REs,\cite{Fulop-2014-80702-Alfvenic} but no clear experimental evidences have been given. In this letter, we present the first experimental observation of TAEs accompanied by the bursts of REs during minor disruptions of purely ohmically heated plasmas without any auxiliary heating in the SUNIST spherical tokamak.

SUNIST is a small spherical tokamak (ST) with major radius $R=0.3$\,m and minor radius $a=0.23$\,m. The experiments discussed here were performed in ohmic plasmas with plasma current $I_{\rm p}=40\,{\rm kA}$, toroidal magnetic field $B_{\rm t}=0.15\,{\rm T}$. The line-averaged density measured by a 94 GHz microwave interferometer was in the range of 0.2-6$\times 10^{18} \,\rm{m}^{-3}$. The hard x-ray (HXR) detected by cadmium-zinc-telluride (CdZnTe) with a narrow collimating aperture was used to estimate the energy of REs.

\begin{figure}
  \includegraphics[width=.75\columnwidth]{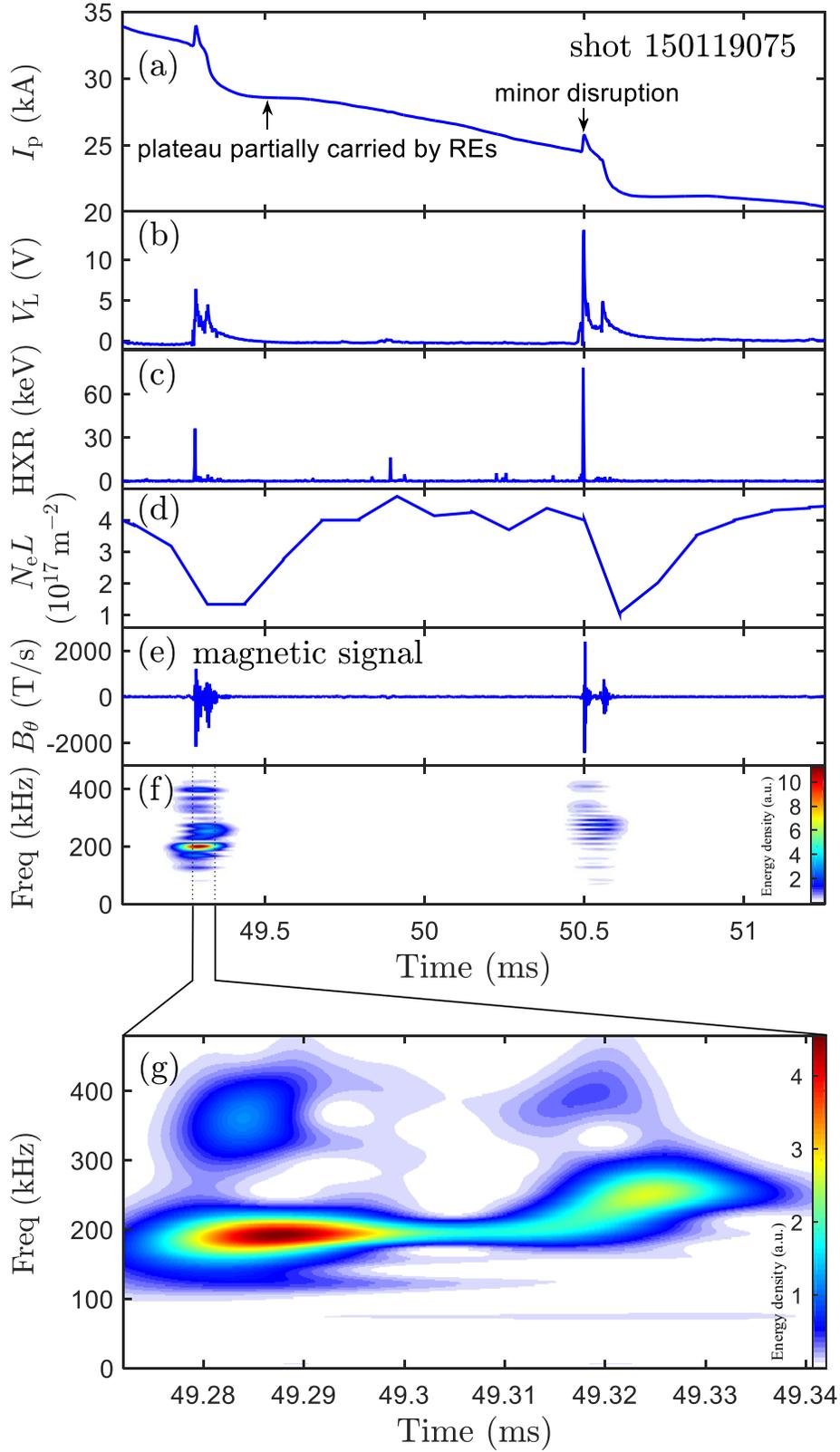} \\ 
  \caption{\label{TAE-time-signals} Time evolution of (a) plasma current, (b) loop voltage, (c) hard x-ray (HXR), (d) line-averaged electron density, (e) magnetic probe signal, (f) the spectrogram of a magnetic probe signal and (g) detail of the wavelet spectrum of a magnetic probe signal in a minor disruption.}
\end{figure}

\begin{figure}
  \includegraphics[width=.9\columnwidth]{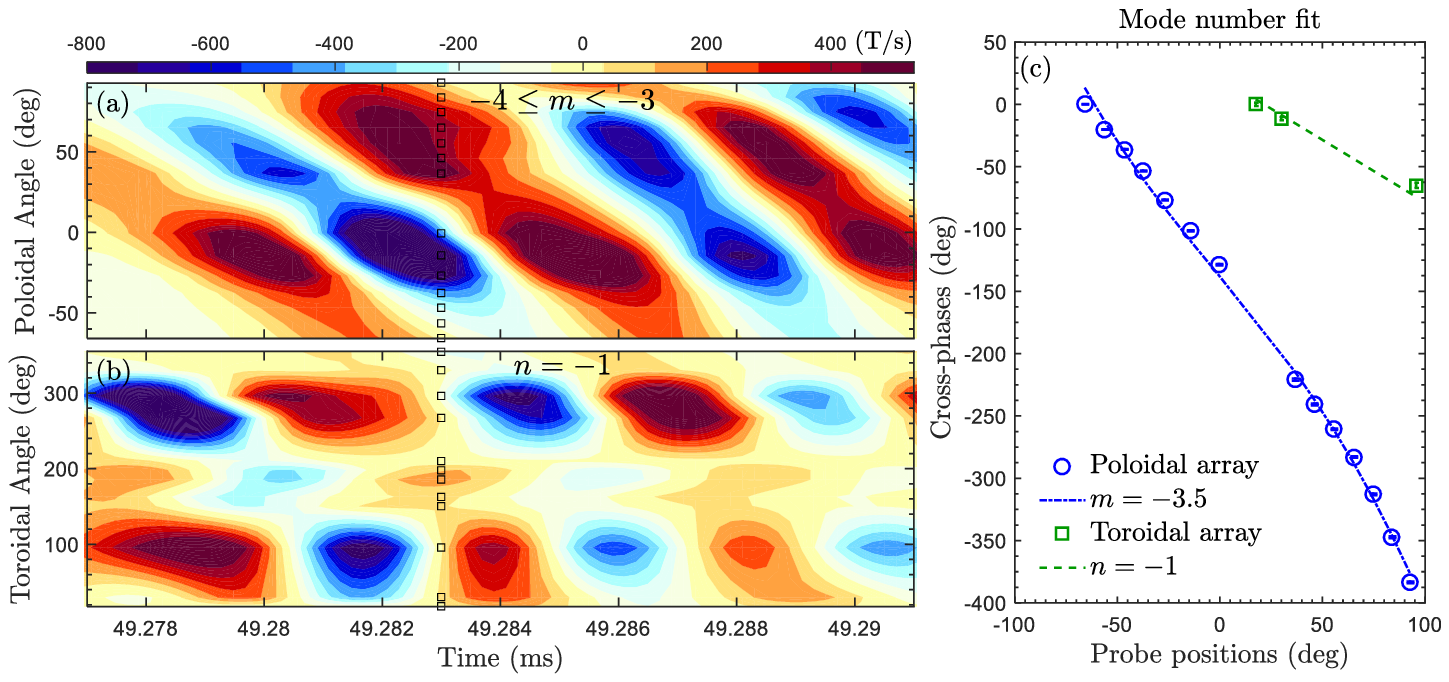} 
  \caption{\label{TAE-mode-number} Left: contour plots of the perturbed high-frequency magnetic field $\delta B_{\theta}$ at TAE frequency range (30-300\,kHz) as a function of time during a minor disruption for shot 150119075. (a) A poloidal array showing that $-4 \leq m \leq -3$ and (b) a toroidal array indicating that $n=-1$ is the dominant spatial mode structure of the field fluctuation. Here, black squares near 49.283\,ms denote magnetic probe locations. Right: (c) the relative phases between all pairs of magnetic signals are plotted as a function of probe position.}
\end{figure}

Owing to the ST configuration, ohmic discharges of SUNIST are seldom terminated by one major disruption. The plasma current often decreases in a stepwise form caused by a sequence of minor disruptions. During these minor disruptions high-frequency magnetic fluctuations are observed from the signals detected by an array of high-frequency magnetic probes (HFMPs) sampled at 15\,MHz.\cite{Liu-2014-11-Design} As shown in Fig. \ref{TAE-time-signals}, a kind of high frequency MHD modes occur during each minor disruptions which all have significant plasma current plateaus partially carried by REs. The frequency range of these modes is 150-500\,kHz. Their toroidal and poloidal mode numbers are $n=-1$ and $-4 \leq m \leq -3$, respectively, as shown in Fig. \ref{TAE-mode-number}. It should be noted that when a phase angle $\xi$ is obtained from each of the HFMPs at the frequency of interest after considering the phase differences between the transfer functions of HFMPs,\cite{Heeter-2000-4092-Fast,Liu-2014-11-Design,Horvath-2015-125005-Reducing} a reliable method for obtaining $m$, $n$ numbers is to find the optimal fit with four free parameters, i.e., $\xi=m(\theta+\lambda\sin\theta)+n\phi+\delta_0$.\cite{Harley-1989-771-TFTR} Here, $\delta_0$ is a phase constant and $\theta$, $\phi$ are the poloidal and toroidal angles of the high-frequency magnetic probes. While $\lambda$ is a free parameter concerning the toroidal effect which enables the fit to allow for toroidal corrections to the geometry of the mode and the off-centre position of the mode with respect to the centre of the vacuum vessel. The mode structure analysis indicates the co-existence of $m/n=-3/-1$ and $-4/-1$ harmonics, propagating toroidally opposite to the direction of plasma current and poloidally in the electron diamagnetic drift direction in the laboratory frame of reference.

\begin{figure}
  \includegraphics[width=.75\columnwidth]{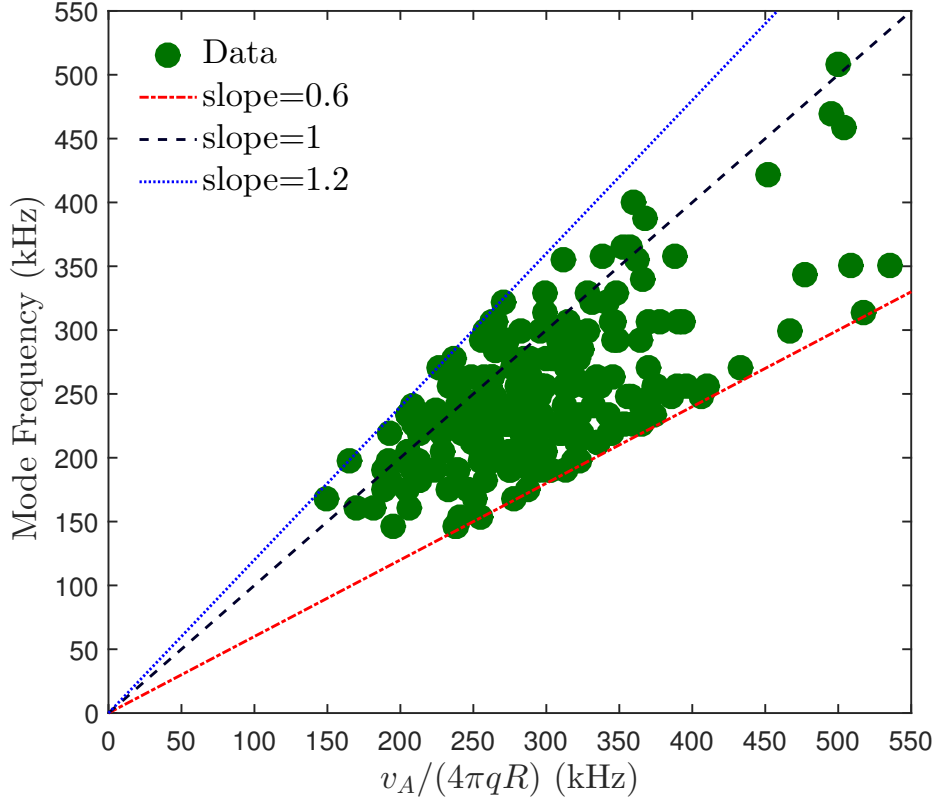} 
  \caption{\label{TAE-freq} Comparison of the observed mode frequencies during minor disruptions with the expected TAE frequencies.}
\end{figure}

There are several minor disruptions in one discharge shown in Fig. \ref{TAE-time-signals}. It can be found that the mode frequency is higher at low density and vice versa, suggesting that the modes may be Alfv\'{e}n-type modes. In order to verify it, a statistical analysis with about 240 independent shots was made. In these shots, minor disruptions and TAEs were almost simultaneously observed. Fig. \ref{TAE-freq} illustrates that the observed mode frequency scales linearly with the TAE frequency $f_{\rm TAE}=v_{\rm A}/4\pi qR$, where $v_{\rm A}=B_{\rm t}/\sqrt{\mu_0 \rho}$ is the Alfv\'{e}n velocity, $\rho$ is the mass density, $q$ is the safety factor and $R$ is the major radius. In the calculations, the line-averaged density and the on-axis toroidal field are used to estimate the Alfv\'{e}n velocity, and the safety factor is estimated by mode numbers. The scatter of data points in Fig. \ref{TAE-freq} is mainly due to the variation of the size of plasma column on which the line-averaged density is depended. Therefore, these high frequency MHD modes should be TAEs.

\begin{figure*}
  \includegraphics[width=.9\textwidth]{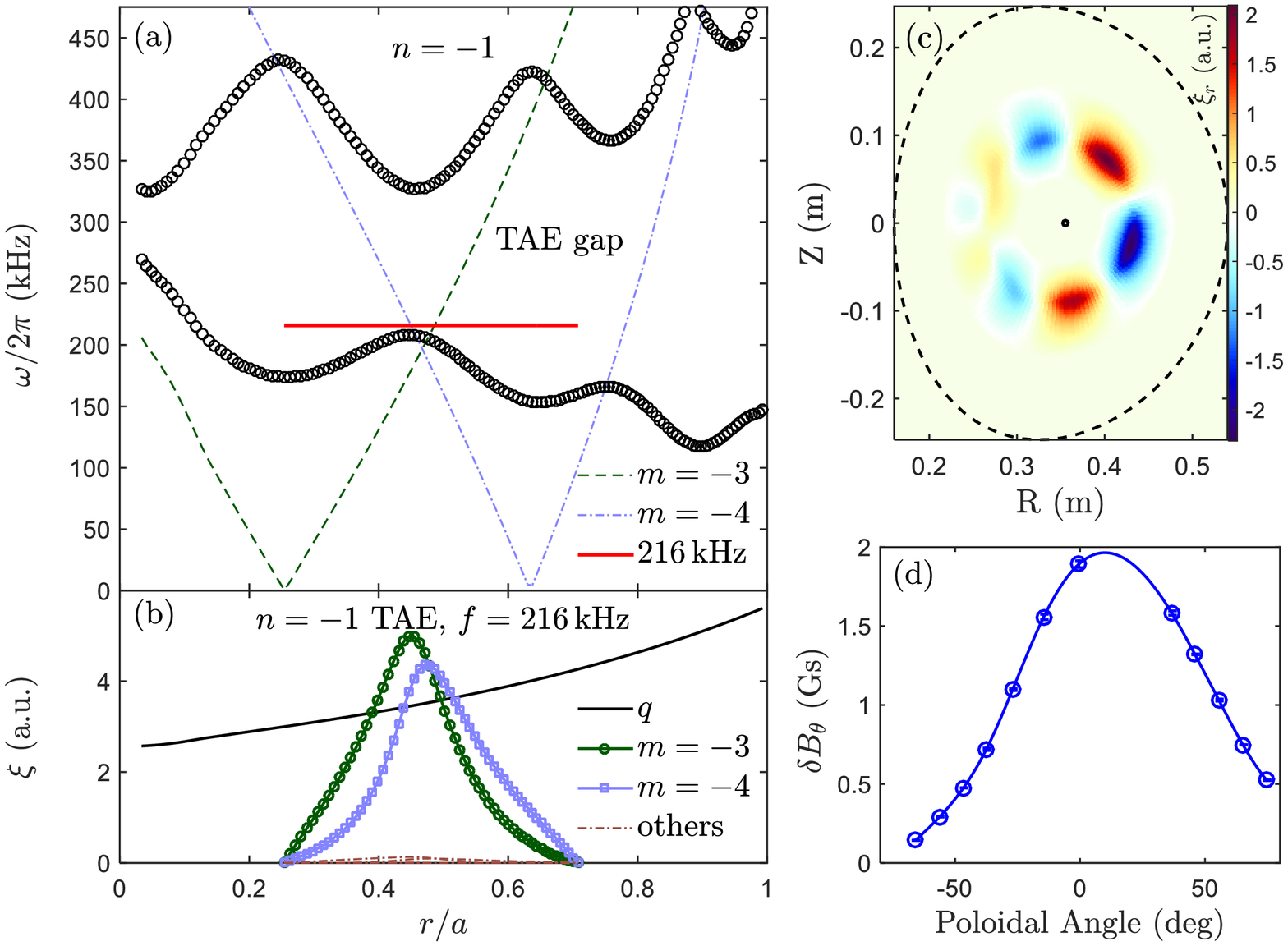} 
  \caption{\label{TAE-mode-structure} (a) The $n=-1$ Alfv\'{e}n continuum calculated with an estimated $q$ profile showing that the 216\,kHz mode lies in the TAE gap. Also plotted are the $m=-3$ and $m=-4$ Alfv\'{e}n continua in the cylindrical geometry limit. The red line denotes the radial range used in the numerical calculation, which is chosen in order to avoid the continuum resonance. (b) Radial profile of the safety factor and radial structure of the amplitude of the $n=-1$ TAE mode at $f=216\,\rm kHz$. (c) Two-dimensional structure of the real part of radial plasma displacement $\xi$ of the $n=-1$ TAE mode. (d) Measurement of the poloidal variation of the mode amplitudes for the TAE. Here, $0^\circ$ is at the outer midplane, and positive angles correspond to the upper part. The equilibrium is for SUNIST shot 150119075 at 49.29\,ms.}
\end{figure*}

The GTAW code\cite{Hu-2014-52510-Numerical} was employed to calculate the frequency and mode structure of Alfv\'{e}n eigenmodes in SUNIST equilibrium. Figure \ref{TAE-mode-structure}(a) shows that a mode with $f=216\,\rm kHz$ is just within the $n=-1$ TAE gap due to the coupling of the $m=-3$ and $m=-4$ harmonics. The radial structure of the mode is plotted in Fig. \ref{TAE-mode-structure}(b), which shows the $m=-3$ and $m=-4$ harmonics are dominant. These dominant harmonics can also be seen in the contour plot of the radial displacement of the mode on the poloidal plane in Fig. \ref{TAE-mode-structure}(c). Moreover, Figure \ref{TAE-mode-structure}(c) shows that this TAE mode exhibits a ballooning-like structure,\cite{Chang-1995-1469-Alfven,Wong-1999-1-review} i.e., the displacement on the low field side is stronger than that on the high field side. This character is consistent with the poloidal variation of measured magnetic field fluctuations, which is given in Fig. \ref{TAE-mode-structure}(d).

TAEs observed in SUNIST are excited at the moment of minor disruptions in pure ohmic regimes, when energetic REs can be generated. Given that fast ions are practically absent in the ohmic plasmas of SUNIST, and the measured mode rotation direction is in the electron direction in the lab frame, energetic REs generated during minor disruptions are considered to be a possible driver of TAEs due to a resonant interaction of REs with the precession drift frequency $\omega_{\rm d}=\frac{Eq}{eB_{\rm t} Rr}\frac{\kappa^2+4r/R}{2r/R+(1-r/R)\kappa^2}g(\kappa)$ for circulating particles ($\kappa<1$),\cite{Zonca-2007-1588-Electron} where $E$ is the energy of the resonant REs, $B_{\rm t}$ is the toroidal field on axis, $R$ is the major radius, $q$ is the safety factor associated with the resonant surface localized at the small radius $r$. Considering $\kappa\sim0$ for passing REs and the experimentally measured TAE frequency in the SUNIST case, the expression for the averaged RE energy that would match the precession drift resonance condition is approximately $E({\rm eV}) \sim 2\pi f_{\rm TAE} B_{\rm t} R r / q$, where $f_{\rm TAE}$ is the observed TAE frequency. Taking, $f_{\rm TAE} \sim 216\,{\rm kHz}, B_{\rm t}=0.15\,{\rm T}, R=0.3\,{\rm m}, r=0.11\,{\rm m}, q=3.5$ for SUNIST shot 150119075 at 49.29\,ms, $E \sim 1.9\,{\rm keV}$. Given the low electron density ($<6\times 10^{18} \,\rm{m}^{-3}$) and the low electron temperature ($\sim 100\,\rm eV$) in SUNIST ohmic plasmas, such $1.9\,{\rm keV}$ electrons can be regarded as REs. Considering the sampling effect of collimated CdZnTe detectors and the broad energy spectrum of REs, the HXR burst shown in Fig. \ref{TAE-time-signals}(c) implies the existence of a large amount of REs with energy in the vicinity of $1.9\,{\rm keV}$. The resonance condition for RE driven TAEs appears to be satisfied for these experimental conditions for SUNIST shot 150119075 accompanying with the change of the electron density. It is noted that the density sweeping during the minor disruption is important for matching the resonant condition. In Fig. \ref{TAE-time-signals}(g), when the density decreases continuously, the TAEs are excited twice, with different frequencies corresponding to different density. On the contrary, although there is another HXR burst with energy 16\,keV for SUNIST shot 150119075 at 49.89\,ms shown in Fig. \ref{TAE-time-signals}(c), the resonance condition for RE driven TAEs may not be satisfied since the electron density is nearly unchanged. Although such mechanism mentioned above appears to be satisfied from the point of RE energy, it is still difficult to understand in physics. After all, that REs drive TAEs through wave-particle resonance in the precession frequency is just a preliminary conjecture. Other possible mechanisms are still open for interpreting the experimental observations.

Recently in the MAST\cite{Gryaznevich-private-communication} and J-TEXT,\cite{Chen-private-communication} there were also observed some high frequency Alfv\'{e}n-like MHD oscillations during minor or major disruptions. All of these indicates there should be some stories between Alfv\'{e}n instabilities and disruption/RE. However, the driving mechanisms of these TAEs are still open. For example, another possible mechanism might be the distortion of the electron distribution function due to the generation of REs near the trapped to passing boundary, where the bounce frequency of electrons can be resonant with TAEs.\cite{Zonca-private-communication}

The experimental data presented here identify TAEs in the ohmic plasmas of SUNIST. These TAEs are observed during minor disruptions with plasma current plateaus partially carried by REs. The measured mode frequency is consistent with the calculated TAE frequency, while the mode propagates in the electron diamagnetic drift direction. From the wave-precession drift resonance mechanism, the excitation condition of TAE is calculated and can be satisfied in the SUNIST spherical tokamak. However, the mechanism of TAEs excitation by runaway electrons is still an open question. More theoretical and experimental efforts are needed to interpret these observations.

The authors sincerely thank Dr. L. Zeng, Dr. W. Chen, Prof. Y.F. Liang, Prof. X.T. Ding and Prof. F. Zonca for helpful discussions and constructive comments. This work was supported by the National Natural Science Foundation of China (Grant Nos. 11325524, 11261140327, 11475102, and 11405218) and the Ministry of Science and Technology of China (Contract Nos. 2013GB112001 and 2013GB107001).

\end{document}